\documentclass[nofootinbib,prd,twocolumn ]{revtex4}%
\usepackage{amsmath}
\usepackage{amsfonts}
\usepackage{amssymb}
\usepackage{graphics}
\usepackage{color}
\usepackage{graphicx}%
\newcommand{\be}{\begin{equation}}
\newcommand{\ee}{\end{equation}}
\newcommand{\beq}{\begin{eqnarray}}
\newcommand{\eeq}{\end{eqnarray}}

\begin{document}
\title{Does Gravity's Rainbow induce Inflation without an Inflaton?}
\author{Remo Garattini}
\email{Remo.Garattini@unibg.it}
\affiliation{Universit\`{a} degli Studi di Bergamo, Facolt\`{a} di Ingegneria,}
\affiliation{Viale Marconi 5, 24044 Dalmine (Bergamo) Italy}
\affiliation{and I.N.F.N. - sezione di Milano, Milan, Italy}
\author{Mairi Sakellariadou}
\email{Mairi.Sakellariadou@kcl.ac.uk}
\affiliation{Department of Physics, King's College London, University of London,}
\affiliation{Strand, London WC2R 2LS, U.K.}

\begin{abstract}
We study aspects of quantum cosmology in the presence of modified space-time
geometry. In particular, within the context of Gravity's Rainbow modified
geometry, motivated from quantum gravity corrections at the Planck energy
scale, we show that the distortion of the metric leads to a Wheeler-De Witt
equation whose solution admits outgoing plane waves. Hence, a period of
cosmological inflation may arise without the need of introducing an inflaton field.

\end{abstract}
\maketitle

\begin{flushleft}
KCL-PH-TH/2012-48
\end{flushleft}

%%%%%%%%%%%%%%%%%%%%%%%%%%%%%%%%%%%%%%%%

\section{Introduction}

The inflationary paradigm~\cite{Starobinsky:1980te} can successfully address
some of the shortcomings of the hot big bang cosmological model and provide a
spectrum of adiabatic fluctuations that can fit the cosmic microwave
background temperature anisotropies data. This widely accepted scenario enrich
the standard cosmological model, combining general relativity with high energy
physics. The \textsl{ conventional} inflationary scenario requires particular
initial conditions for the onset of inflation~\cite{onset} and the
introduction of a scalar field, the inflaton, with a suitable potential so
that a long period of slow-roll inflation can be
accommodated.\footnote{Alternative approaches can be also found in the
literature. For instance, an exponential expansion may be obtained by adding
an $R^{2}$-term in the gravitational action, without changing the particle
content of the theory~\cite{Gottlober:1990um}.} Moreover, cosmic microwave
background temperature anisotropies measurements impose constraints on the
parameters of the inflationary potential. However, the naturalness of a given
inflationary model can be argued only if the inflaton field with the
appropriate potential can result from a fundamental
theory~\cite{Bezrukov:2007ep}. Hence, despite the simplicity of a scalar field
driven slow-roll inflationary era and the promising phenomenological
consequences, in particular regarding the spectrum of cosmic microwave
background temperature anisotropies, this paradigm is not fully satisfactory,
still lacking a solid theoretical framework. Moreover, even though inflation
takes place at energy scales, high enough that quantum corrections can no
longer be neglected, scalar field driven slow-roll inflationary models have
been mainly studied within the classical approximation regime.

At very high energy scales, the simple framework of a smooth manifold with a
metric following classical general relativity breaks down. The threshold is
defined by the Planck energy scale, $E_{\mathrm{Pl}}=\sqrt{\hbar c^{5}/G}$,
above which one should consider a full quantum theory to describe the
structure of space-time. Even though various different approaches (e.g.,
string theory, loop quantum gravity, noncommutative geometry, causal dynamical
triangulations, causal sets) have being developed, we are still lacking a full
quantum gravity theory.

In what follows, we will use the fact that general relativity provides a
natural scheme for quantization of the gravitational field, namely the
Wheeler-De Witt (WDW) equation, which is a quantum version of the Hamiltonian
constraint obtained from the Arnowitt-Deser-Misner decomposition of
space-time. Moreover, we will consider that near the Planck energy scale,
quantum gravity effects may modify the space-time geometry, from the simple
Friedmann-Lema\^{\i}tre-Robertson-Walker (FLRW) metric describing a
homogeneous, isotropic and closed universe with line element
\begin{equation}
ds^{2}=-N^{2}dt^{2}+a^{2}\left(  t\right)  d\Omega_{3}^{2}~, \label{FRW}%
\end{equation}
where $d\Omega_{3}^{2}(k)$ is the metric on the spatial sections which have
constant curvature $k=0,\pm1$, defined by
\begin{equation}
d\Omega_{3}^{2}=\gamma_{ij}dx^{i}dx^{j}~;
\end{equation}
$N=N(t)$ is the lapse function taken to be homogeneous and $a(t)$ denotes the
scale factor. One may for instance consider that near the Planck energy scale,
space-time coordinates may not commute, as in the case of $\kappa$-deformed
Minkowski space-time, hence leading to a modification of the metric.

Let us consider that the space-time geometry is described by the deformed
metric
\begin{equation}
ds^{2}=-\frac{N^{2}\left(  t\right)  }{g_{1}^{2}\left(  E/E_{\mathrm{Pl}%
}\right)  }dt^{2}+\frac{a^{2}\left(  t\right)  }{g_{2}^{2}\left(
E/E_{\mathrm{Pl}}\right)  }d\Omega_{3}^{2}~, \label{FRWMod}%
\end{equation}
where the $g_{1}(E/E_{\mathrm{Pl}})$ and $g_{2}(E/E_{\mathrm{Pl}}) $ functions
of energy, which capture the deformation of the metric in Eq.~$\left(
\ref{FRWMod}\right)  $ above, should be such that
\begin{align}
\lim_{E/E_{\mathrm{Pl}}\rightarrow0}g_{1}\left(  E/E_{\mathrm{Pl}}\right)   &
=1\nonumber\\
\mbox{and}\ \ \ \ \lim_{E/E_{\mathrm{Pl}}\rightarrow0}g_{2}\left(
E/E_{\mathrm{Pl}}\right)   &  =1~,
\end{align}
so that at low energies one recovers the FLRW metric.

It is worth noting that since the $g_{1}$ and $g_{2}$ functions do not depend
on the coordinates, one may absorb them in the lapse function and the scale
factor by redefining an energy dependent time coordinate. This will result to
modifications on the comoving horizon which, depending on the choice of the
$g_{1}(E/E_{\mathrm{Pl}}) $ and $g_{2}(E/E_{\mathrm{Pl}}) $ functions, may
resolve the horizon problem~\cite{MagSmo}. Moreover, in the case of a
spherically symmetric metric, one may absorb the energy dependence of the time
and radial coordinates of the Schwarzschild metric, leading to implications
for black hole thermodynamics~~\cite{MagSmo}. Such proposals remain however
beyond the scope of our present work.

The deformed metric, Eq.~$\left(  \ref{FRWMod}\right)  $, dubbed as gravity's
rainbow has been introduced in Ref.~\cite{MagSmo}, as an extension to the
non-linear, Deformed, or Doubly Special Relativity (DSR)~\cite{DSR} in a
curved space-time. Doubly special relativity modifies special relativity so as
to preserve an additional (apart from the one given by the speed of light)
scale, which corresponds to the scale at which the quantum nature of
space-time reveals itself. In the context of DSR one argues that quantum
gravity effects may alter the expression of the energy and momentum of a
relativistic particle, so that while Lorentz symmetry is preserved, the action
of the Lorentz group becomes non-linear in momentum space, leading to a
consistency with the presence of a new invariant energy/momentum scale related
to the Planck scale. Thus, DSR leads to a modified dispersion relation for a
massive particle.

The deformed metric, Eq.~$\left(  \ref{FRWMod}\right)  $, that we will work
with, may be presented as a generalization of DSR, however one has to keep in
mind that it does not necessarily have to be seen as a proposal intimately
related to DSR. Indeed, if one postulates the validity of gravity's rainbow,
to obtain modified dispersion relations one necessarily has to impose that
plane wave solutions%
\begin{align}
dx^{\mu}p_{\mu} &  =dx^{0}p_{0}+dx^{i}p_{i}\nonumber\\
&  =g^{00}dx_{0}p_{0}+g^{ij}dx_{i}p_{j}~,\label{plane}%
\end{align}
which remain plane, even if the metric is described by Eq.~$\left(
\ref{FRWMod}\right)  $; only in this way can one argue that the modified
dispersion relations are a consequence of gravity's rainbow. However, if we
relax the condition of imposing plane wave solutions even at the Planck scale,
the distorted metric, Eq.~$\left(  \ref{FRWMod}\right)  $, leads to a
Generalized Uncertainty Principle (GUP) which differs from the Heisenberg
uncertainty principle, by terms linear and quadratic in particle momenta. The
GUP induced terms become relevant near the Planck scale and lead to the
existence of a minimum measurable length. This could be interpreted as the
breakdown of the validity of continuum space-time at very small scales.

Our aim here is to study aspects of quantum cosmology in the presence of
gravity's rainbow, which results from a metric deformation as described in
Eq.~(\ref{FRWMod}), through the rainbow functions $g_{1}(E/E_{\mathrm{Pl}}) $
and $g_{2}(E/E_{\mathrm{Pl}})$. Note that the deformation of the geometry is
probed through particles of energy $E$. Hence, $E$ denotes the energy scale at
which quantum gravity effects become apparent. For instance, one may think of
a graviton distorting the FLRW metric at the Planck scale. We will address the
question of whether a successful inflationary era can arise in the absence of
a scalar field (inflaton), as a result of the modified WDW equation in the
context of gravity's rainbow.

In particular, we will investigate whether a closed universe can spontaneously
nucleate out of nothing, and then undergo an inflationary period. In this
approach, the evolution of the universe can be viewed as the motion of a
hypothetical particle having zero energy, coordinates defined by the scale
factor of the universe and moving in a specified potential, whose form leads
to a scale that separates ultra-violet from infra-red physics. Whether the
tunneling of the wave-function that creates a universe out of nothing, can
lead to an inflationary stage or not, will depend on the scale to which the
universe tunnels. In other words, only if there is an exponential expansion in
a short time scale may one expect an inflationary era. We will thus study
whether at large-scales the potential approaches values such that the
wave-function admits outgoing plane waves leading to an expanding universe in
the future. In the affirmative case, we have obtained an inflationary
expansion resulting from the deformation of the metric, without the need to
introduce a scalar field.

The rest of the paper is organized as follows: in Section~\ref{p1}, we recall
the basic elements of quantum cosmology in a FLRW space-time and in
Section~\ref{p2}, we extend our discussion in the presence of gravity's
rainbow. In Section~\ref{p3} we discuss the modifications of the
wave-functions for the \textsl{tunneling} and \textsl{no boundary} proposals
in the context of gravity's rainbow and the onset of inflation. We summarize
our results in Section~\ref{p4}. In Appendix~A we summarize the
\textsl{tunneling} and \textsl{no boundary} proposals. In Appendix~B we
present some technical points. Throughout this analysis we use units in which
$\hbar=c=k=1$.

%%%%%%%%%%%%%%%%%%%%%%%%%%%%%%%%%%%%%%%%%

\section{The WDW equation for the FLRW space-time}

\label{p1}

Let us consider a simple mini-superspace model described by the FLRW line
element, Eq.~(\ref{FRW}). In this background, the Ricci curvature tensor and
the scalar curvature read
\begin{equation}
R_{ij}=\frac{2}{a^{2}\left(  t\right)  }\gamma_{ij}\qquad\mathrm{and}\qquad
R=\frac{6}{a^{2}\left(  t\right)  }~,
\end{equation}
respectively. The Einstein-Hilbert action in $(3+1)$-dim is%
\begin{align}
S &  =\frac{1}{16\pi G}\int_{\Sigma\times I}\mathcal{L}dtd^{3}%
x\nonumber\label{action}\\
&  =\frac{1}{16\pi G}\int_{\Sigma\times I}N\sqrt{g}\left[  K^{ij}K_{ij}%
-K^{2}\right.  \nonumber\\
&  ~~~~~~~~~~~~~~~~~~~~~~~\left.  +R-2\Lambda\right]  dtd^{3}x~,
\end{align}
with $\Lambda$ the cosmological constant, $K_{ij}$ the extrinsic curvature and
$K$ its trace. Using the FLRW line element, Eq.~$\left(  \ref{FRW}\right)  $,
the above action, Eq.~$\left(  \ref{action}\right)  $, becomes
\begin{equation}
S=-\frac{3\pi}{4G}\int_{I}\left[  \dot{a}^{2}a-a+\frac{\Lambda}{3}%
a^{3}\right]  dt~,
\end{equation}
where we have computed the volume associated with the three-sphere,
$V_{3}=2\pi^{2}$, and set $N=1$. The canonical momentum reads
\begin{equation}
\pi_{a}=\frac{\delta S}{\delta\dot{a}}=-\frac{3\pi}{2G}\dot{a}a~,
\end{equation}
and the resulting Hamiltonian density is%
\begin{align}
\mathcal{H} &  =\pi_{a}\dot{a}-\mathcal{L}\nonumber\\
&  =-\frac{G}{3\pi a}\pi_{a}^{2}-\frac{3\pi}{4G}a+\frac{3\pi}{4G}\frac
{\Lambda}{3}a^{3}~.\label{H0}%
\end{align}
Following the canonical quantization prescription, we promote $\pi_{a}$ to a
momentum operator, setting $\pi_{a}\rightarrow-i\partial/\partial a$. Thus,
the WDW equation, $\mathcal{H}\Psi\left(  a\right)  =0$, with $\Psi(a)$ the
wave-function of the universe, reads
\begin{equation}
\left[  -a^{-q}\left[  \frac{\partial}{\partial a}a^{q}\frac{\partial
}{\partial a}\right]  +\frac{9\pi^{2}}{4G^{2}}\left(  a^{2}-\frac{\Lambda}%
{3}a^{4}\right)  \right]  \Psi\left(  a\right)  =0~;\label{WDW_0}%
\end{equation}
for $q=0$ it assumes the familiar form of a one-dimensional Schr\"{o}dinger
equation for a hypothetical particle moving in the potential
\begin{equation}
U\left(  a\right)  =\frac{9\pi^{2}}{4G^{2}}a^{2}\left(  1-\frac{a^{2}}%
{a_{0}^{2}}\right)  ~,\label{U(a)}%
\end{equation}
with zero total energy and coordinate $a(t)$. The parameter $q$ represents the
factor-ordering ambiguity and $a_{0}=\sqrt{3/\Lambda}$ plays the r\^{o}le of a
reference length. The potential, Eq.~(\ref{U(a)}), vanishes at
\begin{equation}
a=a_{0}\qquad\mathrm{or}\qquad a=0~.\nonumber
\end{equation}
For $0<a<a_{0}$, Eq.~$\left(  \ref{U(a)}\right)  $ implies $U\left(  a\right)
>0$, which is the classically forbidden region where the behavior of
$\Psi\left(  a\right)  $ is exponential. For $a>a_{0}$, Eq.~$\left(
\ref{U(a)}\right)  $ leads to $U\left(  a\right)  <0$, which is the
classically allowed region where the behavior of $\Psi\left(  a\right)  $ is oscillatory.

The WDW equation can be solved exactly for the special case of operator
ordering $q=-1$~\cite{Vilenkin}, in terms of Airy functions\footnote{Let us
remind the reader of the asymptotic forms of the Airy functions for large
values of the argument $\left(  z\rightarrow+\infty\right)  $:%
\begin{equation}%
\begin{array}
[c]{c}%
Ai\left(  z\right)  \sim\frac{1}{2\sqrt{\pi}}z^{-\frac{1}{4}}\exp\left(
-\frac{2}{3}z^{\frac{3}{2}}\right)  ~,\nonumber\\
\\
Bi\left(  z\right)  \sim\frac{1}{\sqrt{\pi}}z^{-\frac{1}{4}}\exp\left(
\frac{2}{3}z^{\frac{3}{2}}\right)  ~,\nonumber\\
\\
Ai\left(  -z\right)  \sim\frac{1}{\sqrt{\pi}}z^{-\frac{1}{4}}\sin\left(
\frac{2}{3}z^{\frac{3}{2}}+\frac{\pi}{4}\right)  ~,\nonumber\\
\\
Bi\left(  -z\right)  \sim\frac{1}{\sqrt{\pi}}z^{-\frac{1}{4}}\cos\left(
\frac{2}{3}z^{\frac{3}{2}}+\frac{\pi}{4}\right)  ~.\nonumber
\end{array}
\label{asy}%
\end{equation}
}, as
\begin{align}
\Psi\left(  a\right)   &  =\alpha Ai\left(  z_{0}\left(  1-\frac{a^{2}}%
{a_{0}^{2}}\right)  \right)  \nonumber\\
&  +\beta Bi\left(  z_{0}\left(  1-\frac{a^{2}}{a_{0}^{2}}\right)  \right)
~,\label{Sol}%
\end{align}
where%
\begin{equation}
z_{0}=\left(  \frac{3\pi a_{0}^{2}}{4G}\right)  ^{\frac{2}{3}}.\label{z(a)}%
\end{equation}
For an arbitrary choice of the factor ordering\footnote{There are exact
solutions also for $q=3.$}, the solution of Eq.~(\ref{WDW_0}) in the WKB
approximation acquires, independently of the boundary conditions, a factor
$a^{-(q+1)/2}$~\cite{Wiltshire}. The coefficients $\alpha$ and $\beta$ are
arbitrary complex numbers determined by the choice of the state for the
wave-function of the universe. For example\footnote{See Appendix~A for
details.}, setting $\alpha=0$ and $\beta\neq0$, one obtains the
Vilenkin~(\textsl{tunneling}) wave-function $\Psi_{\mathrm{V}}$%
~\cite{Vilenkin}, while setting $\alpha\neq0$ and $\beta=0$, one obtains the
Hartle-Hawking~(\textsl{no boundary}) wave-function $\Psi_{\mathrm{HH}}%
$~\cite{HH}. It is worth noting that, keeping $\alpha$ and $\beta$
undetermined, one obtains an infinite number of solutions~\cite{LGGG}.

Following Ref.~\cite{LGGG}, we define%
\begin{equation}
\Gamma=\left\vert \frac{\Psi\left(  0\right)  }{\Psi\left(  z_{0}\right)
}\right\vert ^{2}~, \label{Gamma}%
\end{equation}
as the probability coefficient describing the creation of the universe from
\textsl{nothing}; it is an exact probability. If the wave-function is such
that the probability coefficient $\Gamma$ is less than unity, we say that
there is decay or quantum tunneling. Substituting Eq.~$\left(  \ref{Gamma}%
\right)  $ in Eq.~$\left(  \ref{Sol}\right)  $, we obtain
\begin{align}
\Gamma &  =\left\vert \frac{\alpha Ai\left(  0\right)  +\beta Bi\left(
0\right)  }{\alpha Ai\left(  z_{0}\right)  +\beta Bi\left(  z_{0}\right)
}\right\vert ^{2}\nonumber\\
&  =\frac{1}{\left[  \Gamma\left(  2/3\right)  \right]  ^{2}}\sqrt[3]%
{\frac{16\pi^{4}a_{0}^{2}}{G}}\nonumber\\
&  ~~\times\left\vert \frac{\alpha\sqrt{1/3}+\beta}{\alpha\exp\left(
-\frac{2}{3}z_{0}^{\frac{3}{2}}\right)  +2\beta\exp\left(  \frac{2}{3}%
z_{0}^{\frac{3}{2}}\right)  }\right\vert ^{2}~, \label{Gamma1}%
\end{align}
where we have assumed that $a_{0}^{2}\gg G$ and $a^{2}\leq a_{0}^{2}$.

Note that for generic boundary conditions $\Psi\left(  0\right)  $ appears as
a superposition of the under-barrier WKB solutions of Eq.~$\left(
\ref{WDW_0}\right)  $. Namely, disregarding the pre-exponential factor, we
have%
\begin{equation}
\lim_{a\rightarrow0}\Psi_{\pm}\left(  a\right)  =\lim_{a\rightarrow0}%
\exp\left[  \pm\int_{a}^{a_{0}}\left\vert p\left(  a^{\prime}\right)
\right\vert da^{\prime}\right]  ~,
\end{equation}
where%
\begin{equation}
p\left(  a\right)  =\sqrt{U\left(  a\right)  }%
\end{equation}
is the classical momentum and $U(a)$ is the potential given in Eq.~(\ref{U(a)}).

After nucleation, the evolution of the universe becomes classical with initial
value determined by $a^{2}=a_{0}^{2}$. As shown in Appendix~A, given the
wave-functions $\Psi_{\mathrm{V}}$ or $\Psi_{\mathrm{HH}}$, one can easily
compute the probability distribution for the initial values of $a$ in a
nucleating universe. Using the inner product $(\ref{IP})$, the corresponding
conserved current $(\nabla\cdot j=0)$ reads
\begin{equation}
j=\exp\left(  \pm\frac{4}{3}z_{0}^{\frac{3}{2}}\right)  =\exp\left(  \pm
\frac{3\pi}{\Lambda G}\right)  ~,\label{j}%
\end{equation}
in agreement with the semiclassical nucleation probability
\begin{equation}
P\sim\exp\left[  \pm2\int_{0}^{a_{0}}\left\vert p\left(  a^{\prime}\right)
\right\vert da^{\prime}\right]  ~,\label{PrV}%
\end{equation}
where the upper sign $(+)$ corresponds to the wave-function $\Psi
_{\mathrm{HH}}$ and the lower one $(-)$ to $\Psi_{\mathrm{V}}$.

Within the context of inflation, it is assumed that the universe starts out
with a large effective cosmological constant, which arises from the potential
$V\left(  \phi\right)  $ of a scalar field $\phi$, assumed to be homogeneous
and isotropic. The Hamiltonian constraint, Eq.~$\left(  \ref{H0}\right)  $,
reads%
\begin{align}
\mathcal{H}  &  =\pi_{a}\dot{a}+\pi_{\phi}\dot{\phi}-\mathcal{L}\left(
a,\phi\right) \nonumber\label{Inflaton}\\
&  =-\frac{G}{3\pi a}\pi_{a}^{2}-\frac{\pi_{\phi}^{2}}{2a^{3}}+\pi^{2}%
a^{3}V\left(  \phi\right) \nonumber\\
&  \ \ \ \ -\frac{3\pi}{4G}a+\frac{3\pi}{4G}\frac{\Lambda}{3}a^{3}~,
\end{align}
and Eq.~$\left(  \ref{WDW_0}\right)  $ becomes \begin{widetext}\be
\left[  -a^{-q}\left[  \frac{\partial}{\partial a}a^{q}\frac{\partial
}{\partial a}\right]  -\frac{3\pi}{2Ga^{2}}\frac{\partial^{2}}{\partial
\phi^{2}}+\frac{9\pi^{2}}{4G^{2}}\left\{  a^{2}-a^{4}\left(  \frac{\Lambda}
{3}+\frac{4G}{3\pi}V\left(  \phi\right)  \right)  \right\}  \right]
\Psi\left(  a,\phi\right)  =0~,\label{WDWap}%
\ee
\end{widetext}
where the field $\phi$ can be regarded as a parameter. Assuming that $V\left(
\phi\right)  $ and $\Psi\left(  a,\phi\right)  $ are slowly varying functions
of $\phi$, we can neglect derivatives with respect to $\phi$. Then the WDW
equation, Eq.~$\left(  \ref{WDWap}\right)  $, reduces to%
\begin{equation}
\left[  -\frac{\partial^{2}}{\partial a^{2}}-\frac{q}{a}\frac{\partial
}{\partial a}+U\left(  a,\phi\right)  \right]  \Psi\left(  a,\phi\right)  =0~,
\label{WDWapr}%
\end{equation}
where%
\begin{equation}
U\left(  a,\phi\right)  =\frac{9\pi^{2}}{4G^{2}}a^{2}\left(  1-\frac{a^{2}%
}{a_{\mathrm{eff}}^{2}}\right)  ~ \label{U(a,phi)}%
\end{equation}
and%
\begin{align}
a_{\mathrm{eff}}^{2}  &  ={\frac{3}{\Lambda_{\mathrm{eff}}(\phi)}}\nonumber\\
&  =a_{0}^{2}\left[  1+\frac{4G}{\pi\Lambda}V\left(  \phi\right)  \right]
^{-1}~. \label{aeff}%
\end{align}
Equation~$\left(  \ref{WDWapr}\right)  $ can be solved, for $q=-1$, in terms
of Airy's functions. The effect of the inflaton is the modification of the
\textsl{reference length} $a_{0}$, since the cosmological constant $\Lambda$
is replaced by an effective cosmological constant with the additional
contribution of the potential of the scalar field.

An attempt to avoid the introduction of the inflaton field has been proposed
in Ref.~\cite{WPKA}, where a time-dependent cosmological constant has been
considered. In the next section we will study the effect of a distortion of
the FLRW space-time metric to examine whether the inflaton can be substituted
by a pure gravitational field with quantum fluctuations.

%%%%%%%%%%%%%%%%%%%%%%%%%%%%%%%%%%%%%%%%%%

\section{The WDW equation in the context of Gravity's Rainbow}

\label{p2}

Let us use the distortion of the metric produced by gravity's rainbow, namely
substitute the line element Eq.~$\left(  \ref{FRW}\right)  $ with Eq.~$\left(
\ref{FRWMod}\right)  $. The form of the background is such that the shift
function $N^{i}$ vanishes. The extrinsic curvature reads
\begin{align}
K_{ij} &  =-\frac{\dot{g}_{ij}}{2N}\nonumber\\
&  =\frac{g_{1}\left(  E/E_{\mathrm{Pl}}\right)  }{g_{2}^{2}\left(
E/E_{\mathrm{Pl}}\right)  }\tilde{K}_{ij}~,\label{Kij}%
\end{align}
where an overdot denotes differentiation with respect to time. The trace of
the extrinsic curvature is
\begin{equation}
K=g_{1}\left(  E/E_{\mathrm{Pl}}\right)  \tilde{K}%
\end{equation}
and the momentum conjugate to the three-metric $g_{ij}$ of $\Sigma$ is
\begin{align}
\pi^{ij} &  =\frac{\sqrt{g}}{16\pi G}\left(  Kg^{ij}-K^{ij}\right)
\nonumber\\
&  =\frac{g_{1}\left(  E/E_{\mathrm{Pl}}\right)  }{g_{2}\left(
E/E_{\mathrm{Pl}}\right)  }\tilde{\pi}^{ij}~.
\end{align}
Note that any quantity with a tilde indicates the same quantity computed in
the absence of gravity's rainbow functions $g_{1}\left(  E/E_{\mathrm{Pl}%
}\right)  $ and $g_{2}\left(  E/E_{\mathrm{Pl}}\right)  $.

The distorted classical constraint, $\mathcal{H}=0$, then
reads\begin{widetext}
\begin{equation}
16\pi G\frac{g_{1}^{2}\left(  E/E_{\mathrm{Pl}}\right)  }{g_{2}^{3}\left(
E/E_{\mathrm{Pl}}\right)  }\tilde{G}_{ijkl}\tilde{\pi}^{ij}\tilde{\pi}^{kl}
\mathcal{-}\frac{\sqrt{\tilde{g}}}{16\pi Gg_{2}\left(  E/E_{\mathrm{Pl}
}\right)  }\!{}\!\left[  \tilde{R}-\frac{2\Lambda}{g_{2}^{2}\left(
E/E_{\mathrm{Pl}}\right)  }\right]   =0~, \label{Acca}
\end{equation}
\end{widetext}where we have used that
\begin{equation}
R=g_{2}^{2}\left(  E/E_{\mathrm{Pl}}\right)  \tilde{R}~,\label{RR}%
\end{equation}
and%
\begin{align}
G_{ijkl} &  =\frac{1}{2\sqrt{g}}\left(  g_{ik}g_{jl}+g_{il}g_{jk}-g_{ij}%
g_{kl}\right)  \nonumber\\
&  =\frac{\tilde{G}_{ijkl}}{g_{2}\left(  E/E_{\mathrm{Pl}}\right)  }~.
\end{align}
From Eq.~$\left(  \ref{Acca}\right)  $, one can easily see that the distorted
WDW equation $\mathcal{H}\Psi=0$ becomes\begin{widetext}
\begin{equation}
\left[  16\pi G\frac{g_{1}^{2}\left(  E/E_{\mathrm{Pl}}\right)  }{g_{2}%
^{3}\left(  E/E_{\mathrm{Pl}}\right)  }\tilde{G}_{ijkl}\tilde{\pi}^{ij}%
\tilde{\pi}^{kl}  \mathcal{-}\frac{\sqrt{\tilde{g}}}{16\pi Gg_{2}\left(
E/E_{\mathrm{Pl}}\right)  }\!{}\!\left(  \tilde{R}-\frac{2\Lambda}{g_{2}%
^{2}\left(  E/E_{\mathrm{Pl}}\right)  }\right)  \right]  \Psi(a)
=0~.\label{AccaR}%
\end{equation}
\end{widetext}Integrating out all degrees of freedom except the scale factor,
one gets\begin{widetext}
\begin{equation}
\left[  -\frac{G}{3\pi a}\pi_{a}^{2}\frac{g_{1}^{2}\left(  E/E_{\mathrm{Pl}%
}\right)  }{g_{2}^{3}\left(  E/E_{\mathrm{Pl}}\right)  }-\frac{3\pi}{4G}%
\frac{a}{g_{2}\left(  E/E_{\mathrm{Pl}}\right)  }
+\frac{\pi}{4G}\frac{a^{3}\Lambda}{g_{2}^{3}\left(  E/E_{\mathrm{Pl}%
}\right)  }\right]  \Psi\left(  a\right)    =0~. \label{WDWg0}%
\end{equation}
\end{widetext}Assuming that the factor ordering is not distorted by the
presence of the gravity's rainbow functions, one can further simplify the
above equation and get%
\begin{equation}
\left[  -\frac{\partial^{2}}{\partial a^{2}}-\frac{q}{a}\frac{\partial
}{\partial a}+U\left(  a,E/E_{\mathrm{Pl}}\right)  \right]  \Psi\left(
a\right)  =0,\label{WDWg}%
\end{equation}
where we have set $N=1$ and defined the distorted potential as%
\begin{align}
U\left(  a,E/E_{\mathrm{Pl}}\right)   &  =\left[  \frac{3\pi g_{2}\left(
E/E_{\mathrm{Pl}}\right)  }{2Gg_{1}\left(  E/E_{\mathrm{Pl}}\right)  }\right]
^{2}\nonumber\\
&  \ \ \ \times a^{2}\left[  1-\frac{a^{2}}{a_{0}^{2}g_{2}^{2}\left(
E/E_{\mathrm{Pl}}\right)  }\right]  ~,\label{U(a,E)}%
\end{align}
with $a_{0}=\sqrt{3/\Lambda}$. The distorted potential above, reduces to the
potential given in Eq.$\left(  \ref{U(a)}\right)  $ in the limit of
$E\rightarrow0$. A remark on Eqs.$\left(  \mathbf{\ref{WDWg}}\right)  $ and
$\left(  \ref{U(a,E)}\right)  $ is in order. One can think that when
$g_{1}\left(  E/E_{\mathrm{Pl}}\right)  =C_{1}$ and $g_{2}\left(
E/E_{\mathrm{Pl}}\right)  =C_{2}$ with $C_{1}$ and $C_{2}$ constants, the
original WDW equation $\left(  \ref{WDW_0}\right)  $ still manifests a
distortion due to Gravity's Rainbow. However, this is not the case, because by
simply rescaling $G$ and $\Lambda$ in the following way%
\begin{equation}
G\rightarrow G^{\prime}=G\frac{g_{1}\left(  E/E_{\mathrm{Pl}}\right)  }%
{g_{2}\left(  E/E_{\mathrm{Pl}}\right)  }\qquad\mathrm{and}\qquad
\Lambda\rightarrow\Lambda^{\prime}=\frac{\Lambda}{g_{2}^{2}\left(
E/E_{\mathrm{Pl}}\right)  },
\end{equation}
Eq.$\left(  \ref{WDWg}\right)  $ is formally equivalent to $\left(
\ref{WDW_0}\right)  $ and therefore no effect of Gravity's Rainbow is
measurable.\textbf{ }Let us note that we consider the energy $E$ as a
parameter denoting the energy scale, which is the same for all observers, at
which quantum gravity effects cannot be neglected. Thus, as we have already
discussed, the deformation of space-time, modeled through the rainbow
functions $g_{1}\left(  E/E_{\mathrm{Pl}}\right)  $ and $g_{2}\left(
E/E_{\mathrm{Pl}}\right)  $ is probed through particles of energy $E$.
Comparing the distorted potential, Eq.$\left(  \ref{U(a,E)}\right)  $, with
the potential containing the inflaton, Eq.$\left(  \ref{U(a,phi)}\right)  $,
we conclude that the rainbow functions may play the r\^{o}le of the inflaton
field. Hence, $g_{1}\left(  E/E_{\mathrm{Pl}}\right)  $, $g_{2}\left(
E/E_{\mathrm{Pl}}\right)  $ and $E$ can be regarded as parameters. Finally,
let us emphasize that since we are considering quantum gravity effects,
resulting in a space-time modification, it is consistent not to work with the
classical constraint, Eq.$\left(  \ref{Acca}\right)  $, but to use the WDW
equation, Eq.$\left(  \ref{AccaR}\right)  $, instead.

Coming back to the distorted potential, Eq.$\left(  \ref{U(a,E)}\right)  $, we
see that it vanishes when%
\begin{equation}
a=g_{2}\left(  E/E_{\mathrm{Pl}}\right)  a_{0}\qquad\mathrm{or}\qquad
a=0~,\nonumber
\end{equation}
and gets its maximum%
\begin{equation}
U_{\mathrm{max}}=\left[  \frac{3\pi g_{2}^{2}\left(  E/E_{\mathrm{Pl}}\right)
a_{0}}{4Gg_{1}\left(  E/E_{\mathrm{Pl}}\right)  }\right]  ^{2}~,
\end{equation}
when%
\begin{equation}
a=g_{2}\left(  E/E_{\mathrm{Pl}}\right)  \frac{a_{0}}{\sqrt{2}}~.\nonumber
\end{equation}
For $0<a<g_{2}\left(  E/E_{\mathrm{Pl}}\right)  a_{0}$, the potential
$U\left(  a,E/E_{\mathrm{Pl}}\right)  $ is positive, leading to a classically
forbidden (Euclidean) region with an exponential wave-function $\Psi\left(
a,E/E_{\mathrm{Pl}}\right)  $, while for $a>g_{2}\left(  E/E_{\mathrm{Pl}%
}\right)  a_{0}$, the potential is negative, leading to a classically allowed
(Lorentzian) region where the behavior of $\Psi\left(  a,E/E_{\mathrm{Pl}%
}\right)  $ is oscillatory. The height and shape of $U\left(
a,E/E_{\mathrm{Pl}}\right)  $ depend strongly on the choice of $g_{1}\left(
E/E_{\mathrm{Pl}}\right)  $ and $g_{2}\left(  E/E_{\mathrm{Pl}}\right)  $. The
WDW equation for gravity's rainbow has the conventional form and only the
potential is affected by the introduction of the functions $g_{1}%
(E/E_{\mathrm{Pl}})$ and $g_{2}(E/E_{\mathrm{Pl}})$ which modify the line
element. Thus, the solution of the WDW equation Eq.~$\left(  \ref{WDWg}%
\right)  $ with the operator ordering $q=-1$ is
\begin{align}
\Psi\left(  a\right)   &  =\alpha Ai\left(  z_{0}\left(  1-\frac{a^{2}}%
{a_{0}^{2}g_{2}^{2}\left(  E/E_{\mathrm{Pl}}\right)  }\right)  \right)
\nonumber\\
&  \ \ \ +\beta Bi\left(  z_{0}\left(  1-\frac{a^{2}}{a_{0}^{2}g_{2}%
^{2}\left(  E/E_{\mathrm{Pl}}\right)  }\right)  \right)  ~,\label{SolMod}%
\end{align}
where $z_{0}$ depends on $E$ and is given by
\begin{equation}
z_{0}(E/E_{\mathrm{Pl}})=\left[  \frac{3\pi a_{0}^{2}g_{2}^{3}\left(
E/E_{\mathrm{Pl}}\right)  }{4Gg_{1}\left(  E/E_{\mathrm{Pl}}\right)  }\right]
^{\frac{2}{3}}~.\label{z(a)Mod}%
\end{equation}
In particular, $g_{2}\left(  E/E_{\mathrm{Pl}}\right)  $ sets up the
nucleation point and the nucleation probability, while $g_{1}\left(
E/E_{\mathrm{Pl}}\right)  $ is related only to the nucleation probability.
Indeed, following the same steps that have led to Eqs.~$\left(  \ref{Gamma}%
\right)  $, $\left(  \ref{Gamma1}\right)  $, we get that the probability
coefficient $\Gamma$ becomes%
\begin{align}
\Gamma &  =\left\vert \frac{\Psi\left(  g_{2}\left(  E/E_{\mathrm{Pl}}\right)
a_{0}\right)  }{\Psi\left(  0\right)  }\right\vert ^{2}\nonumber\\
&  =\frac{\pi g_{2}^{3}\left(  E/E_{\mathrm{Pl}}\right)  }{\sqrt{3}\left[
\Gamma\left(  2/3\right)  \right]  ^{2}}\sqrt[3]{\frac{\pi a_{0}^{2}}{4G}%
}\left\vert \frac{\alpha+\beta\sqrt{3}}{\alpha\exp\left(  -\mathcal{A}\right)
+2\beta\exp\mathcal{A}}\right\vert ^{2}\label{GammaR}%
\end{align}
where we have defined
\begin{equation}
\mathcal{A}\equiv\frac{\pi a_{0}^{2}g_{2}^{3}\left(  E/E_{\mathrm{Pl}}\right)
}{2Gg_{1}\left(  E/E_{\mathrm{Pl}}\right)  }~,
\end{equation}
and we have assumed that $a_{0}^{2}\gg G$ and $a^{2}\leq\left[  a_{0}%
g_{2}\left(  E/E_{\mathrm{Pl}}\right)  \right]  ^{2}$. Since the nucleation
point is at $a=g_{2}\left(  E/E_{\mathrm{Pl}}\right)  a_{0}$ and $a_{0}$ is
the observed radius of the universe, we need to impose that $g_{2}\left(
E/E_{\mathrm{Pl}}\right)  a_{0}\ll1$ when the energy scale is too high. This
condition is required to get a small tunneling barrier; otherwise one would
get nucleation in today's universe. In addition, this choice is in agreement
with the fact that $a_{0}g_{2}\left(  E/E_{\mathrm{Pl}}\right)  $ plays the
r\^{o}le of $1/\sqrt{V\left(  \phi\right)  }$; at the beginning of inflation
the potential of the inflaton must be large. Moreover, inflation ends when
$g_{2}\left(  E/E_{\mathrm{Pl}}\right)  =1$, which is compatible with the low
energy limit of the rainbow's functions.

Let us examine the conditions on the relation between the rainbow functions so
that there is quantum tunneling. If $g_{1}(E/E_{\mathrm{Pl}})$ vanishes faster
than $g_{2}^{3}(E/E_{\mathrm{Pl}})$, then%
\begin{equation}
\exp\left(  \pm\mathcal{A}\right)  \rightarrow\left\{
\begin{array}
[c]{c}%
+\infty\text{ for the plus sign}\\
\\
0\text{ for the minus sign}%
\end{array}
\right.  \label{faster}%
\end{equation}
and%
\begin{equation}
\Gamma\simeq\frac{\pi g_{2}^{3}\left(  E/E_{\mathrm{Pl}}\right)  }{2\beta
\sqrt{3}\left[  \Gamma\left(  2/3\right)  \right]  ^{2}}\sqrt[3]{\frac{\pi
a_{0}^{2}}{4G}}\left\vert \left(  \alpha+\beta\sqrt{3}\right)  \exp\left(
-\mathcal{A}\right)  \right\vert ^{2}~. \label{eq-Gamma}%
\end{equation}
Note that Eq.~$\left(  \ref{eq-Gamma}\right)  $ above is strictly speaking
only valid for the Vilenkin boundary conditions. For the Hartle-Hawking
boundary conditions, one will have first to set $\beta=0$ and then discuss the
limits on $g_{1}(E/E_{\mathrm{Pl}})$ and $g_{2}(E/E_{\mathrm{Pl}})$. Since it
turns out that $\Gamma<1$, one gets that both $\Psi_{\mathrm{V}}$ and
$\Psi_{\mathrm{HH}}$ wave-functions predict tunneling even in the presence of
gravity's rainbow.\footnote{If one considers the rainbow's gravity as a
generalization of the DSR, then since for massless particles the modified
dispersion relation that results from DSR cannot fix the r\^{o}le of
$g_{1}\left(  E/E_{\mathrm{Pl}}\right)  $ and $g_{2}\left(  E/E_{\mathrm{Pl}%
}\right)  $, the choices made in Refs.~\cite{AM,GaMaLo,GaMa} are not in
conflict with the ones made above, Eq.~$\left(  \ref{faster}\right)  $ here.
\newline}

In the opposite case, namely if $g_{1}(E/E_{\mathrm{Pl}})$ vanishes slower
than $g_{2}^{3}(E/E_{\mathrm{Pl}})$, then
\begin{equation}
\exp\left(  \pm\mathcal{A}\right)  \rightarrow1\label{slower}%
\end{equation}
for both plus/minus signs, which implies that $\Gamma<1$, leading to quantum
tunneling only if the condition
\begin{equation}
\frac{\pi g_{2}^{3}\left(  E/E_{\mathrm{Pl}}\right)  }{\sqrt{3}\left[
\Gamma\left(  2/3\right)  \right]  ^{2}}\sqrt[3]{\frac{\pi a_{0}^{2}}{4G}}<1
\end{equation}
is satisfied.

%%%%%%%%%%%%%%%%%%%%%%%%%%%%%%%%%%

\section{The wave-function of the Universe in the presence of Gravity's
Rainbow and the Onset of Inflation}

\label{p3}

We will discuss how the wave-function of the universe, within the
Hartle-Hawking and Vilenkin proposals, is modified in the presence of
gravity's rainbow. The under-barrier wave-functions $\Psi_{\mathrm{V}}$ and
$\Psi_{\mathrm{HH}}$ become \begin{widetext}
\begin{equation}
\Psi_{\mathrm{V}}^{\mathrm{inside}}\left(  a\right)   \simeq\left(
1-\frac{a^{2}}{g_{2}^{2}\left(  E/E_{\mathrm{Pl}}\right)  a_{0}^{2}}\right)
^{-\frac{1}{4}}\exp\left[  -\frac{2}{3}z_{0}^{\frac{3}{2}}\left(  E/E_{\mathrm{Pl}
}\right)  \left\{  1-\left(  1-\frac{a^{2}}{g_{2}^{2}\left(  E/E_{\mathrm{Pl}
}\right)  a_{0}^{2}}\right)  ^{\frac{3}{2}}\right\}  \right]~,\nonumber
\end{equation}
\end{widetext}
and \begin{widetext}
\begin{equation}
\Psi_{\mathrm{HH}}^{\mathrm{inside}}\left(  a\right)    \simeq\left(
1-\frac{a^{2}}{g_{2}^{2}\left(  E/E_{\mathrm{Pl}}\right)  a_{0}^{2}}\right)
^{-\frac{1}{4}}
\exp\left[  \frac{2}{3}z_{0}^{\frac{3}{2}}\left(  E/E_{\mathrm{Pl}%
}\right)  \left\{  1-\left(  1-\frac{a^{2}}{g_{2}^{2}\left(  E/E_{\mathrm{Pl}%
}\right)  a_{0}^{2}}\right)  ^{\frac{3}{2}}\right\}  \right]~,\nonumber
\end{equation}
\end{widetext}
respectively. In this case, $\Psi\left(  0\right)  $ appears as a
superposition of the under-barrier WKB solutions of Eq.~$\left(
\ref{WDWg}\right)  $, namely%
\begin{equation}
\lim_{a\rightarrow0}\Psi_{\pm}\left(  a\right)  =\lim_{a\rightarrow0}%
\exp\left[  \pm\int_{a}^{g_{2}\left(  E/E_{\mathrm{Pl}}\right)  a_{0}}p\left(
a^{\prime}\right)  da^{\prime}\right]  ~,
\end{equation}
where%
\begin{equation}
p\left(  a\right)  =\left[  U\left(  a,E/E_{\mathrm{Pl}}\right)  \right]
^{\frac{1}{2}}%
\end{equation}
is the classical momentum, with $U(a,E/E_{\mathrm{Pl}})$ the potential given
in Eq.~$\left(  \ref{U(a,E)}\right)  $.

Note that every consideration we have done \textsl{hitherto} is in the
under-barrier region, where trans-Planckian physics is meaningful. Outside the
barrier, we face inevitably classical expansion and therefore the issue of the
onset of inflation is of relevance.

In the Vilenkin proposal, the wave-function reads \begin{widetext}\be
\Psi_{\mathrm{V}}^{\mathrm{outside}}\left(  a\right)  \simeq\left(
\frac{a^{2}}{g_{2}^{2}\left(  E/E_{\mathrm{Pl}}\right)  a_{0}^{2}}-1\right)
^{-\frac{1}{4}}\exp\left[  \left(  -\frac{2}{3}z_{0}^{\frac{3}{2}}\left(
E/E_{\mathrm{Pl}}\right)  \right)  \left(  1+i\left(  \frac{a^{2}}{g_{2}%
^{2}\left(  E/E_{\mathrm{Pl}}\right)  a_{0}^{2}}-1\right)  ^{\frac{3}{2}%
}\right)  +i\frac{\pi}{4}\right] ~,\label{PsiVO}%
\ee
\end{widetext}and for the Hartle-Hawking proposal, the wave-function
is\begin{widetext}
\be
\Psi_{\mathrm{HH}}^{\mathrm{outside}}  \simeq2\left(  \frac{a^{2}}%
{g_{2}^{2}\left(  E/E_{\mathrm{Pl}}\right)  a_{0}^{2}}-1\right)  ^{-\frac
{1}{4}}\exp\left(  \frac{2}{3}z_{0}^{\frac{3}{2}}\left(  E/E_{\mathrm{Pl}%
}\right)  \right)  \cos\left(  \frac{2}{3}z^{\frac{3}{2}}-\frac{\pi}%
{4}\right)~.\label{PsiHHO}%
\ee
\end{widetext}To show that there is an inflating solution, we recall that
independently of the boundary conditions outside the barrier, we have that for
large scale factors%
\begin{equation}
\frac{a^{2}}{g_{2}^{2}\left(  E/E_{\mathrm{Pl}}\right)  a_{0}^{2}}%
\gg1~,\nonumber
\end{equation}
and for
\begin{equation}
g_{2}\left(  E/E_{\mathrm{Pl}}\right)  \sim1~,\nonumber
\end{equation}
the wave function takes the form%
\begin{equation}
\Psi^{\mathrm{out}}\left(  a\right)  \simeq\exp\left(  iS\right)  \simeq
\exp\left[  \pm i\left(  \frac{a^{2}}{a_{0}^{2}}-1\right)  ^{\frac{3}{2}%
}\right]  ~.\label{psi-out}%
\end{equation}
In the above, Eq.~(\ref{psi-out}), we have neglected the pre-factor and $S$
satisfies the Hamilton-Jacobi equation%
\begin{equation}
\left[  \left(  \frac{\partial S}{\partial a}\right)  ^{2}+U\left(
a,E/E_{\mathrm{Pl}}\right)  \right]  =0.\label{HJ}%
\end{equation}
Comparison with Eq.$\left(  \ref{WDWg0}\right)  $ suggests that we identify%
\begin{equation}
\frac{\partial S}{\partial a}=\pi_{a}=-\frac{3\pi}{2G}\dot{a}a,
\end{equation}
which implies%
\begin{align}
\frac{\dot{a}}{a} &  =\left[  \frac{g_{2}\left(  E/E_{\mathrm{Pl}}\right)
}{g_{1}\left(  E/E_{\mathrm{Pl}}\right)  }\right]  \sqrt{\frac{1}{a_{0}%
^{2}g_{2}^{2}\left(  E/E_{\mathrm{Pl}}\right)  }-\frac{1}{a^{2}}}~\nonumber\\
&  \simeq\sqrt{\frac{\Lambda}{3}}\frac{1}{g_{1}\left(  E/E_{\mathrm{Pl}%
}\right)  },\label{infl}%
\end{align}
where we have assumed that $a^{2}\gg a_{0}^{2}g_{2}^{2}\left(
E/E_{\mathrm{Pl}}\right)  $. Equation $\left(  \ref{infl}\right)  $
corresponds to an inflationary solution if $E$ does not depend on time $t$,
otherwise one has to consider the specific form of $E\left(  t\right)  $ and
$g_{1}\left(  E/E_{\mathrm{Pl}}\right)  $. However, this goes beyond the
purpose of this paper. It is interesting to note that, to first order in
$E/E_{\mathrm{Pl}}$ one gets%
\begin{equation}
\frac{\dot{a}}{a}=\sqrt{\frac{\Lambda}{3}}~\left(  1-g_{1}^{^{\prime}}\left(
0\right)  \frac{E}{E_{\mathrm{Pl}}}\right)
\end{equation}
and the corresponding inflationary solution is slowed or accelerated depending
on the sign and therefore on the form of $g_{1}\left(  E/E_{\mathrm{Pl}%
}\right)  $. Next step is to show that there is sufficient inflation. To do
this we compute the conserved current for the Vilenkin and the Hartle-Hawking
proposals, $j_{\mathrm{V}}$ and $j_{\mathrm{HH}}$, respectively. It is easy to
show that they are in agreement with the respective nucleation probabilities:
\begin{align*}
j_{\mathrm{V}}(E) &  =\exp\left(  -\frac{4}{3}z_{0}^{\frac{3}{2}}\right)
=\exp\left(  -\frac{\pi a_{0}^{2}g_{2}^{3}\left(  E/E_{\mathrm{Pl}}\right)
}{Gg_{1}\left(  E/E_{\mathrm{Pl}}\right)  }\right)  \\
&  =P_{\mathrm{V}}\sim\exp\left[  -2\int_{0}^{g_{2}\left(  E/E_{\mathrm{Pl}%
}\right)  a_{0}}\left\vert p\left(  a^{\prime}\right)  \right\vert da^{\prime
}\right]
\end{align*}
and%
\begin{align*}
j_{\mathrm{HH}}(E) &  =\exp\left(  \frac{4}{3}z_{0}^{\frac{3}{2}}\right)
=\exp\left(  \frac{\pi g_{2}^{3}\left(  E/E_{\mathrm{Pl}}\right)  a_{0}^{2}%
}{Gg_{1}\left(  E/E_{\mathrm{Pl}}\right)  }\right)  \\
&  =P_{\mathrm{HH}}\sim\exp\left[  2\int_{0}^{g_{2}\left(  E/E_{\mathrm{Pl}%
}\right)  a_{0}}\left\vert p\left(  a^{\prime}\right)  \right\vert da^{\prime
}\right]  ~.
\end{align*}
Therefore, the transition between the forbidden and the classical regions
happens in a continuous way. However, outside the barrier we expect that
$(E/E_{\mathrm{Pl}})\ll1$; thus the distortion on the currents due to
gravity's rainbow is%
\begin{equation}
j_{\mathrm{V}}(E)\simeq\exp\left[  -\frac{\pi a_{0}^{2}}{G}\left(  3{g_{{2}%
}^{\prime}\left(  0\right)  }-{g_{{1}}^{\prime}\left(  0\right)  }\right)
\frac{E}{E_{\mathrm{Pl}}}\right]  \label{jv}%
\end{equation}
for the Vilenkin proposal, and%
\begin{equation}
j_{\mathrm{HH}}(E)\simeq\exp\left[  \frac{\pi a_{0}^{2}}{G}\left(  3{g_{{2}%
}^{\prime}\left(  0\right)  }-{g_{{1}}^{\prime}\left(  0\right)  }\right)
\frac{E}{E_{\mathrm{Pl}}}\right]  ~,\label{jhh}%
\end{equation}
for the Hartle-Hawking proposal, where we have used that $g_{1}\left(
0\right)  =g_{2}\left(  0\right)  =1$.

To discuss the probability to have sufficient inflation we compute conditional
probabilities using the wave-function outside the barrier, namely%
\begin{equation}
P\left(  \phi_{i}>\phi_{\mathrm{suff}}|\ 0<\phi_{i}<\phi_{\sup}\right)
=\frac{\int_{\phi_{\mathrm{suff}}}^{\phi_{\sup}}j\left(  \phi\right)  d\phi
}{\int_{0}^{\phi_{\sup}}j\left(  \phi\right)  d\phi}~, \label{CProb}%
\end{equation}
where $j\left(  \phi\right)  $ is the density current $j=\exp(\pm
4/3z_{0}^{3/2})$, with $z_{0}$ defined in Eq.~$\left(  \ref{z(a)}\right)  $,
$\phi_{\mathrm{suff}}$ denoting the minimum value of $\phi$ which guarantees
$N_{\min}\simeq60$ e-folds of inflation and $\phi_{\sup}$ being a cut-off at
the Planck scale, or even the trans-Planckian one~\cite{CM}. The initial value
is found at $a^{2}=1/V\left(  \phi\right)  $, which is the surface separating
the tunneling from the inflating region.

Following Eq.~$\left(  \ref{CProb}\right)  $ for gravity's rainbow one has%
\begin{equation}
P\left(  E>E_{\mathrm{suff}}|\ 0<E<\infty\right)  =\frac{\int
_{E_{\mathrm{suff}}}^{\infty}j\left(  E\right)  dE}{\int_{0}^{\infty}j\left(
E\right)  dE}~, \label{P(E)}%
\end{equation}
where $j\left(  E\right)  $ is the density current and $E_{\mathrm{suff}}$ is
the minimum value of $E$ for which there are at least 60 e-folds. For
gravity's rainbow, the surface separating the tunneling and inflating region
is located at $a^{2}=a_{0}^{2}g_{2}^{2}\left(  E/E_{\mathrm{Pl}}\right)  $.
Due to the presence of the rainbow's functions $g_{1}\left(  E/E_{\mathrm{Pl}%
}\right)  $ and $g_{2}\left(  E/E_{\mathrm{Pl}}\right)  $, we are allowed to
take the upper limit of the integration to infinity.

For the Vilenkin proposal, we thus obtain \begin{widetext}
\begin{equation}
P_{\mathrm{V}}\left(  E>E_{\mathrm{suff}}|\ 0<E<\infty\right)  =\frac
{\int_{E_{\mathrm{suff}}}^{\infty}\exp\left(  -\frac{\pi a_{0}^{2}g_{2}%
^{3}\left(  E/E_{\mathrm{Pl}}\right)  }{Gg_{1}\left(  E/E_{\mathrm{Pl}%
}\right)  }\right)  dE}{\int_{0}^{\infty}\exp\left(  -\frac{\pi a_{0}^{2}%
g_{2}^{3}\left(  E/E_{\mathrm{Pl}}\right)  }{Gg_{1}\left(  E/E_{\mathrm{Pl}%
}\right)  }\right)  dE}~=1-\frac{\int_{0}^{E_{\mathrm{suff}}}\exp\left(
-\frac{\pi a_{0}^{2}g_{2}^{3}\left(  E/E_{\mathrm{Pl}}\right)  }{Gg_{1}\left(
E/E_{\mathrm{Pl}}\right)  }\right)  dE}{\int_{0}^{\infty}\exp\left(
-\frac{\pi a_{0}^{2}g_{2}^{3}\left(  E/E_{\mathrm{Pl}}\right)  }{Gg_{1}\left(
E/E_{\mathrm{Pl}}\right)  }\right)  dE}~. \label{PgV}%
\end{equation}
\end{widetext}
Let us consider some forms of $g_{1}\left(  E/E_{\mathrm{Pl}}\right)  $ and
$g_{2}\left(  E/E_{\mathrm{Pl}}\right)  $ as discussed in Ref.~\cite{AM}. From
the above probability, Eq.~$\left(  \ref{PgV}\right)  $, it is sufficient to
fix the form of the ratio in the argument of the exponential. Note that we are
forced to consider $g_{2}\left(  E/E_{\mathrm{Pl}}\right)  \ll1$ when $E\gg
E_{\mathrm{Pl}}$ to avoid nucleation in today's universe. In order to have a
finite result, $g_{1}\left(  E/E_{\mathrm{Pl}}\right)  \rightarrow0$ much
faster than $g_{2}\left(  E/E_{\mathrm{Pl}}\right)  \ll1$ when $E\gg
E_{\mathrm{Pl}}$. We will examine some simple cases:

\noindent$\bullet$ For%
\begin{equation}
\frac{g_{2}^{3}\left(  E/E_{\mathrm{Pl}}\right)  }{g_{1}\left(
E/E_{\mathrm{Pl}}\right)  }=1+\alpha\frac{E}{E_{\mathrm{Pl}}}~, \label{1)}%
\end{equation}
with $\alpha>0$, we obtain%
\begin{equation}
P_{\mathrm{V}}=\exp\left(  -3\pi\alpha\frac{E_{\mathrm{suff}}}{\Lambda
GE_{\mathrm{Pl}}}\right)  ~.
\end{equation}
Hence, in this case our proposal can only work if the parameter $\alpha$ is
fine tuned to a very small number. Indeed, if $\Lambda G\sim10^{-120}$ (in
Planck units), to have at least $E_{\mathrm{suff}}\simeq10^{60}E_{\mathrm{Pl}%
}$, we need $\alpha\ll10^{-180}$. Note that the choice $E_{\mathrm{suff}%
}\rightarrow0$ would imply that inflation is sufficient for keV or eV energy,
which is unphysical and therefore discarded.

\noindent$\bullet$ For%
\begin{equation}
\frac{g_{2}^{3}\left(  E/E_{\mathrm{Pl}}\right)  }{g_{1}\left(
E/E_{\mathrm{Pl}}\right)  }=1-\beta\frac{E}{E_{\mathrm{Pl}}}+\alpha\frac
{E^{2}}{E_{\mathrm{Pl}}^{2}}~, \label{2)}%
\end{equation}
with $\alpha,\beta>0$, using the integrals (\ref{Num}) and (\ref{Den}) of
Appendix~\ref{B}, we obtain%
\begin{equation}
P_{\mathrm{V}}=1-\frac{\operatorname{erf}{\left(  {\frac{\sqrt{c}\beta}%
{2\sqrt{\alpha}}}\right)  }+\operatorname{erf}{\left(  {\frac{\sqrt{c}\left(
2\,\alpha\,b-\beta\right)  }{2\sqrt{\alpha}}}\right)  }}{\operatorname{erf}%
{\left(  {\frac{\sqrt{c}\beta}{2\sqrt{\alpha}}}\right)  }+1}~,
\end{equation}
where
\begin{equation}
b=\frac{E_{\mathrm{suff}}}{E_{\mathrm{Pl}}},\qquad c=\frac{3\pi}{\Lambda G}
\label{bc}%
\end{equation}
and $\operatorname{erf}(x)$ is the error function. Note that for
$b\rightarrow0$ the probability $P_{\mathrm{V}}\rightarrow1$, but this case is
disregarded as explained below Eq.~(\ref{1)}) above. Since $c\gg1$ and
$\operatorname{erf}(x)=-\operatorname{erf}(-x)$, we find
\begin{equation}
P_{\mathrm{V}}\rightarrow\left\{
\begin{array}
[c]{cc}%
0 & \ \mathrm{when\qquad}2\alpha b>\beta\\
& \\
1 & \ \mathrm{when\qquad}2\alpha b<\beta
\end{array}
\right.  ~.
\end{equation}
The condition for obtaining sufficient inflation $(E_{\mathrm{suff}%
}/E_{\mathrm{Pl}}\simeq10^{60})$ reads
\begin{equation}
10^{60}<\frac{\beta}{2\alpha}~.
\end{equation}
\noindent$\bullet$ For%
\begin{equation}
\frac{g_{2}^{3}\left(  E/E_{\mathrm{Pl}}\right)  }{g_{1}\left(
E/E_{\mathrm{Pl}}\right)  }=\left\{
\begin{array}
[c]{cc}%
\frac{1}{1+E/E_{\mathrm{Pl}}}\ ,\  & 0\leq E\leq E_{\mathrm{suff}}\\
& \\
-\beta\frac{E}{E_{\mathrm{Pl}}}+\alpha\frac{E^{2}}{E_{\mathrm{Pl}}^{2}}\ ,\  &
E_{\mathrm{suff}}\leq E\leq\infty
\end{array}
\right.  ~, \label{3)}%
\end{equation}
with $\alpha,\beta>0$, we obtain \begin{widetext}
\begin{equation}
P_{\mathrm{V}}=\frac{\int_{E_{\mathrm{suff}}}^{+\infty}\exp\left(  -\frac{\pi
a_{0}^{2}}{G}\left(  -\beta E/E_{\mathrm{Pl}}+\alpha E^{2}/E_{\mathrm{Pl}}%
^{2}\right)  \right)  dE}{\int_{0}^{E_{\mathrm{suff}}}\exp\left(  -\frac{\pi
a_{0}^{2}}{G\left(  1+E/E_{\mathrm{Pl}}\right)  }\right)  dE+\int
_{E_{\mathrm{suff}}}^{+\infty}\exp\left(  -\frac{\pi a_{0}^{2}}{G}\left(
-\beta E/E_{\mathrm{Pl}}+\alpha E^{2}/E_{\mathrm{Pl}}^{2}\right)  \right)
dE}~.\label{Pv1}%
\end{equation}
\end{widetext}Using the integrals of Appendix~\ref{B}, the asymptotic
expression of $P_{\mathrm{V}}$ is%
\begin{equation}
P_{\mathrm{V}}=\frac{\exp\left(  cb(\beta-\alpha b)\right)  }{2c\alpha b^{2}}~
\end{equation}
with $b,c$ defined in Eq.~(\ref{bc}). Clearly,
\begin{equation}
P_{\mathrm{V}}\rightarrow1\qquad\text{\textrm{for\qquad}}\beta=\alpha
b+\frac{\ln\left(  2c\alpha b^{2}\right)  }{cb}.
\end{equation}
Note that the choices proposed in $\left(  \ref{1)}\right)  $,$\left(
\ref{2)}\right)  $ and $\left(  \ref{3)}\right)  $ are not the result of a
series expansion, but they are simply a polynomial truncated at the second
order. For the no boundary proposal the probability density has a positive
argument and choices $\left(  \ref{2)}\right)  $ and $\left(  \ref{3)}\right)
$ lead always to a divergent result. This happens because the integration over
the whole region makes it difficult to assign a meaning for the Hartle-Hawking
state. To consider cases $\left(  \ref{2)}\right)  $ and $\left(
\ref{3)}\right)  $ for the no boundary state, one should map the parameters
$\alpha,\beta$ as
\begin{equation}
\underset{\mathrm{Vilenkin}}{\underbrace{\left(  \alpha,\beta\right)  }}%
\qquad\longrightarrow\qquad\underset{\mathrm{Hartle-Hawking}}{\underbrace
{\left(  -\left\vert \alpha\right\vert ,-\left\vert \beta\right\vert \right)
}}\quad~. \label{VHH}%
\end{equation}
Then, for case $\left(  \ref{2)}\right)  $ with the substitution (\ref{VHH}),
we get $P_{\mathrm{V}}=P_{\mathrm{HH}}$. However, for case $\left(
\ref{3)}\right)  $ only the leading term in $b$ leads to $P_{\mathrm{V}%
}=P_{\mathrm{HH}}$; the subleading terms lead to a complex probability
$P_{\mathrm{HH}}$ (see the asymptotic expansion $(\ref{I1a})$ in Appendix~B)
and therefore one cannot deduce the emergence of an inflationary state.

\section{Conclusions}

\label{p4} Suggesting that our universe underwent, during the early stages of
its evolution, a very fast (exponential) expansion, dubbed as cosmological
inflation, driven by a (usually scalar) field (the inflaton), has become the
standard paradigm. However, there remains the difficulty of finding a
theoretical framework that incorporates the existence of an inflaton field
with the required properties in order to get a successful inflationary model.

In this study, we have tried to pursue another line of approach which is
encoded in general relativity, namely the WDW equation distorted by gravity's
rainbow. The appealing point of Gravity's Rainbow is that it switches on at
the Planck scale, distorting only the trans-Planckian region. Note that this
property has been used to keep under control ultra-violet
divergences~\cite{GaMaLo} and several cosmological applications have been also
considered in the literature (e.g., Ref.~\cite{AM}).

In particular, we have applied gravity's rainbow within the context of quantum
cosmology. We have shown that the rainbow functions $g_{1}\left(
E/E_{\mathrm{Pl}}\right)  $ and $g_{2}\left(  E/E_{\mathrm{Pl}}\right)  $,
which deform the space-time geometry, seem to play the same r\^{o}le as the
inflaton field, at least formally. Whether one will get a successful
inflationary era depends on the choice of the rainbow functions, thus in the
deformation of space-time at the quantum gravity regime. Hence, one may be
able to avoid the need to introduce an external scalar field (the inflaton); a
result which would imply that physics is encoded in the gravitational field
alone. Of course to enforce this point one could also try to implement some
measurement of some quantity, for instance%
\begin{equation}
\left\langle a\right\rangle =\frac{\left\langle \Psi_{\mathrm{HH}}\left\vert
a\right\vert \Psi_{\mathrm{HH}}\right\rangle }{\left\langle \Psi_{\mathrm{HH}%
}|\Psi_{\mathrm{HH}}\right\rangle }\qquad\mathrm{or}\qquad\left\langle
a\right\rangle =\frac{\left\langle \Psi_{\mathrm{V}}\left\vert a\right\vert
\Psi_{\mathrm{V}}\right\rangle }{\left\langle \Psi_{\mathrm{V}}|\Psi
_{\mathrm{V}}\right\rangle }.
\end{equation}
However, the average value of the scale factor needs the knowledge of a ground
state and that is not the purpose of this paper. The same situation happens
for other observables.

%%%%%%%%%%%%%%%%%%%%%%%%%%%%%%%%%%%%%%%%%%%%%%%%%%%%%%%%%%%%%%

\appendix{}

\section{Tunneling and No Boundary Proposals}

\label{A}

To discuss the different boundary conditions we will consider the
under-barrier and the outside-barrier regions separately. Keeping $\alpha$ and
$\beta$ undetermined as in Eq.~$\left(  \ref{Gamma}\right)  $ we find for the
under-barrier region that \begin{widetext}
\be
\frac{\Psi(a)}{\Psi(0)}=\left(  1-\frac{a^{2}}{a_{0}^{2}}\right)  ^{-\frac
{1}{4}}\left[  \frac{\alpha\exp\left(  -\frac{2}{3}z_{0}^{\frac{3}{2}}\left(
1-\frac{a^{2}}{a_{0}^{2}}\right)  ^{\frac{3}{2}}\right)  +2\beta\exp\left(
\frac{2}{3}z_{0}^{\frac{3}{2}}\left(  1-\frac{a^{2}}{a_{0}^{2}}\right)
^{\frac{3}{2}}\right)  }{\alpha\exp\left(  -\frac{2}{3}z_{0}^{\frac{3}{2}%
}\right)  +2\beta\exp\left(  \frac{2}{3}z_{0}^{\frac{3}{2}}\right)  }\right]
~, \label{In}%
\ee
\end{widetext}where we have used the asymptotic approximation of the Airy's
functions with positive arguments. Note that the expression of the probability
coefficient $\Gamma$ in Eq.~$\left(  \ref{Gamma}\right)  $ is exact and
therefore in proximity of the barrier we have no divergences. Outside the
barrier one can use the asymptotic form of the Airy functions for negative
arguments rearranged as \begin{widetext}
\be
\frac{\Psi(a)}{\Psi(0)}\simeq\left(  \frac{a^{2}}{a_{0}^{2}}-1\right)
^{-\frac{1}{4}}
\left[  \frac{\left(  \frac{\alpha}{i}+\beta\right)
\exp\left(  i\frac{2}{3}z_{0}^{\frac{3}{2}}\left(  \frac{a^{2}}{a_{0}^{2}%
}-1\right)  ^{\frac{3}{2}}+i\frac{\pi}{4}\right)  -\left(  \frac{\alpha}%
{i}-\beta\right)  \exp\left(  -i\frac{2}{3}z_{0}^{\frac{3}{2}}\left(
\frac{a^{2}}{a_{0}^{2}}-1\right)  ^{\frac{3}{2}}-i\frac{\pi}{4}\right)
}{\alpha\exp\left(  -\frac{2}{3}z_{0}^{\frac{3}{2}}\right)  +2\beta\exp\left(
\frac{2}{3}z_{0}^{\frac{3}{2}}\right)  }\right]  ~, \label{Out}%
\ee
\end{widetext}where, we have maintained the same normalization as in
Eq.~$\left(  \ref{In}\right)  $. To obtain a unique solution, one has to
specify the boundary conditions. There are two main boundary conditions
proposals, the Vilenkin (\textsl{tunneling}) proposal and the Hartle-Hawking
(\textsl{no boundary}), with wave-functions $\Psi_{\mathrm{V}}$ and
$\Psi_{\mathrm{HH}}$, respectively. In the former, the wave-function
$\Phi(h,\phi)$, with $h_{ij}$ the three-geometry, and $\phi$ the field matter
field configuration,is obtained by integrating over Lorentzian histories
interpolating between a vanishing three-geometry $0$ and $(h,\phi)$ and lying
to the past of$(h,\phi)$; in the latter, the wave-function is given by a
Euclidean path integral over compact four-geometries $g_{\mu\nu}$ bounded by
$h_{ij}$ with field configuration $\phi(x)$. Note that with the choice
Eq.~$\left(  \ref{Gamma}\right)  $, we are forced to require that both
wave-functions $\Psi_{\mathrm{V}}$ and $\Psi_{\mathrm{HH}}$ are bounded as
$a\rightarrow0$. Therefore, we fix
\begin{equation}
\Psi_{\mathrm{HH}}(a)_{|a=0}=\Psi_{\mathrm{V}}(a)_{|a=0}=1.
\end{equation}
For the Vilenkin proposal, the wave-function $\Psi_{\mathrm{V}}$ is defined by
the choice
\begin{equation}
\alpha=0\ \ \mbox{and}\ \ \beta=i~,
\end{equation}
leading to an outgoing wave outside the potential barrier. Note that in
Eq.~$\left(  \ref{In}\right)  $ above, before fixing $\alpha$ and $\beta$, one
has
\begin{equation}
\lim_{z_{0}\rightarrow+\infty}\lim_{a\rightarrow a_{0}}\frac{\Psi(a)}{\Psi
(0)}\neq\lim_{a\rightarrow a_{0}}\lim_{z_{0}\rightarrow+\infty}\frac{\Psi
(a)}{\Psi(0)}~,
\end{equation}
hence only the right hand side limit can lead to the Vilenkin boundary
condition. Moreover, to obtain continuity between Eq.~$\left(  \ref{In}%
\right)  $ and Eq.~$\left(  \ref{Out}\right)  $ at $a=a_{0}$, one has to take
the third root of $\left(  -1\right)  ^{-\frac{1}{4}}$; in agreement with the
WKB approximation of the outgoing and ingoing waves. Let us mention that in
Ref.~\cite{Vilenkin}, the variable $z$ of Eq.~$\left(  \ref{z(a)}\right)  $
was defined as $-z$, thus one takes the principal root of $\left(  -1\right)
^{-\frac{1}{4}}$. Hence, the final form of the wave-function reads
\begin{align}
\Psi_{\mathrm{V}}^{\mathrm{outside}}\left(  a\right)   &  \simeq\left(
\frac{a^{2}}{a_{0}^{2}}-1\right)  ^{-\frac{1}{4}}\nonumber\\
&  \times\exp\left[  \left(  -\frac{2}{3}z_{0}^{\frac{3}{2}}\right)  \left(
1+i\left(  \frac{a^{2}}{a_{0}^{2}}-1\right)  ^{\frac{3}{2}}\right)
+i\frac{\pi}{4}\right]  ~.
\end{align}
For the under barrier wave-function, we simply have
\begin{align}
\Psi_{\mathrm{V}}^{\mathrm{inside}}\left(  a\right)   &  \simeq\left(
1-\frac{a^{2}}{a_{0}^{2}}\right)  ^{-\frac{1}{4}}\nonumber\\
&  \times\exp\left[  \left(  -\frac{2}{3}z_{0}^{\frac{3}{2}}\right)  \left(
1-\left(  1-\frac{a^{2}}{a_{0}^{2}}\right)  ^{\frac{3}{2}}\right)  \right]  ~.
\end{align}
For the Hartle-Hawking proposal, the wave-function $\Psi_{\mathrm{HH}}$ is
defined as
\begin{equation}
\alpha=1\ \ \mbox{and}\ \ \beta=0~.
\end{equation}
Outside the barrier the wave-function reads
\begin{align}
\Psi_{\mathrm{HH}}^{\mathrm{outside}} &  \simeq2\left(  \frac{a^{2}}{a_{0}%
^{2}}-1\right)  ^{-\frac{1}{4}}\nonumber\\
&  \times\exp\left(  \frac{2}{3}z_{0}^{\frac{3}{2}}\right)  \cos\left(
\frac{2}{3}z^{\frac{3}{2}}-\frac{\pi}{4}\right)  ~,
\end{align}
while in the under barrier classically forbidden region
\begin{align}
\Psi_{\mathrm{HH}}^{\mathrm{inside}}\left(  a\right)   &  \simeq\left(
1-\frac{a^{2}}{a_{0}^{2}}\right)  ^{-\frac{1}{4}}\nonumber\\
&  \times\exp\left[  \left(  \frac{2}{3}z_{0}^{\frac{3}{2}}\right)  \left(
1-\left(  1-\frac{a^{2}}{a_{0}^{2}}\right)  ^{\frac{3}{2}}\right)  \right]  ~.
\end{align}
For an arbitrary choice of the factor ordering, the wave-function,
independently of the boundary conditions, acquires a factor $a^{-\left(
q+1\right)  /2}~$~\cite{Wiltshire}. With the wave-functions $\Psi_{\mathrm{V}%
}$ or $\Psi_{\mathrm{HH}}$, one can compute the probability distribution for
the initial values of $a$ in a nucleating universe. The conserved current for
a FLRW mini-superspace reads
\begin{equation}
j=\frac{i}{2}a^{q}\left(  \Psi^{\ast}\partial_{a}\Psi-\Psi\partial_{a}%
\Psi^{\ast}\right)  ~.\label{IP}%
\end{equation}
It is easy to show that for $\Psi_{\mathrm{V}}$ the conserved current is
\begin{equation}
j_{\mathrm{V}}=\exp\left(  -\frac{4}{3}z_{0}^{\frac{3}{2}}\right)
=\exp\left(  -\frac{3\pi}{\Lambda G}\right)  ~,
\end{equation}
in agreement with the nucleation probability $P=\left\vert \frac{\Psi\left(
a_{0}\right)  }{\Psi\left(  0\right)  }\right\vert ^{2}$:
\begin{equation}
P_{\mathrm{V}}\sim\exp\left[  -2\int_{0}^{a_{0}}\left\vert p\left(  a^{\prime
}\right)  \right\vert da^{\prime}\right]  ~,
\end{equation}
where the WKB form of the Vilenkin wave-function $\Psi_{\mathrm{V}}$ has been
used. For the Hartle-Hawking wave-function $\Psi_{\mathrm{HH}}$, the
corresponding conserved current is
\begin{equation}
j_{\mathrm{HH}}=\exp\left(  \frac{4}{3}z_{0}^{\frac{3}{2}}\right)
=\exp\left(  \frac{3\pi}{\Lambda G}\right)  ~,
\end{equation}
in agreement with the tunneling probability
\begin{equation}
P_{\mathrm{HH}}\sim\exp\left[  2\int_{0}^{a_{0}}\left\vert p\left(  a^{\prime
}\right)  \right\vert da^{\prime}\right]  ~,
\end{equation}
where
\begin{equation}
p\left(  a\right)  =\left[  -U\left(  a\right)  \right]  ^{\frac{1}{2}%
}\nonumber
\end{equation}
is the classical momentum, with $U(a)$ the potential given in Eq.~$\left(
\ref{U(a)}\right)  $. Since the observation supports small values of the
cosmological constant, we conclude that in the Vilenkin proposal, we have a
suppression of the nucleation process, while in the Hartle-Hawking proposal
there is exactly the reversed situation.

\section{Integrals}

\label{B}

Let us compute the integrals related to the probability of having sufficient
inflation within the Vilenkin proposal. We begin with the case defined by
Eq.~$\left(  \ref{2)}\right)  $. We have to compute the second term of the
r.h.s. of the probability, namely%
\begin{equation}
\frac{\int_{0}^{E_{\mathrm{suff}}}\exp\left(  -\frac{\pi a_{0}^{2}g_{2}%
^{3}\left(  E/E_{\mathrm{Pl}}\right)  }{Gg_{1}\left(  E/E_{\mathrm{Pl}%
}\right)  }\right)  dE}{\int_{0}^{\infty}\exp\left(  -\frac{\pi a_{0}^{2}%
g_{2}^{3}\left(  E/E_{\mathrm{Pl}}\right)  }{Gg_{1}\left(  E/E_{\mathrm{Pl}%
}\right)  }\right)  dE}. \label{RPV}%
\end{equation}
With the help of the relationship $\left(  \ref{bc}\right)  $, the numerator
of $\left(  \ref{RPV}\right)  $ becomes%
\begin{align}
&  E_{\mathrm{Pl}}\int_{0}^{b}\exp\left(  -c\left(  1-\beta x+\alpha
x^{2}\right)  \right)  dx\nonumber\\
&  =E_{\mathrm{Pl}}\sqrt{\frac{\pi}{4c\alpha}}\exp\left(  c\left(
\frac{{\beta}^{2}}{4\alpha}-1\right)  \right)  \times\nonumber\\
&  \left[  \operatorname{erf}{\left(  {\frac{\sqrt{c}\beta}{2\sqrt{\alpha}}%
}\right)  }+\operatorname{erf}{\left(  {\frac{\sqrt{c}\left(  2\alpha
b-\beta\right)  }{2\sqrt{\alpha}}}\right)  }\right]  \label{Num}%
\end{align}
where $\operatorname{erf}\left(  x\right)  $ is the error function. When
$b\rightarrow\infty$, the corresponding denominator is%
\begin{align}
&  E_{\mathrm{Pl}}\int_{0}^{\infty}\exp\left(  -c\left(  1-\beta x+\alpha
x^{2}\right)  \right)  dx\nonumber\\
&  =E_{\mathrm{Pl}}\sqrt{\frac{\pi}{4c\alpha}}\exp\left(  c\left(
\frac{{\beta}^{2}}{4\alpha}-1\right)  \right)  {^{\,}}\left(
\operatorname{erf}{\left(  {\frac{\sqrt{c}\beta}{2\sqrt{\alpha}}}\right)
}+1\right)  \label{Den}%
\end{align}
and the quantity $\left(  \ref{RPV}\right)  $ simplifies to%
\begin{align}
&  \frac{\int_{0}^{E_{\mathrm{suff}}}\exp\left(  -\frac{\pi a_{0}^{2}g_{2}%
^{3}\left(  E/E_{\mathrm{Pl}}\right)  }{Gg_{1}\left(  E/E_{\mathrm{Pl}%
}\right)  }\right)  dE}{\int_{0}^{\infty}\exp\left(  -\frac{\pi a_{0}^{2}%
g_{2}^{3}\left(  E/E_{\mathrm{Pl}}\right)  }{Gg_{1}\left(  E/E_{\mathrm{Pl}%
}\right)  }\right)  dE}\nonumber\\
&  =\frac{\operatorname{erf}{\left(  {\frac{\sqrt{c}\beta}{2\sqrt{\alpha}}%
}\right)  }+\operatorname{erf}{\left(  {\frac{\sqrt{c}\left(  2\,\alpha
\,b-\beta\right)  }{2\sqrt{\alpha}}}\right)  }}{\operatorname{erf}{\left(
{\frac{\sqrt{c}\beta}{2\sqrt{\alpha}}}\right)  }+1}. \label{Rap}%
\end{align}
For the case defined by Eq.~$\left(  \ref{3)}\right)  $, the first integral in
the denominator of the r.h.s. of Eq.$\left(  \ref{Pv1}\right)  $ becomes%
\begin{align}
&  E_{\mathrm{Pl}}\int_{0}^{b}\exp\left(  -\frac{c}{1+x}\right)  dx\nonumber\\
&  =-\exp\left(  c\right)  +\mathit{E}_{1}\left(  c\right)  c+\exp\left(
-\frac{c}{b+1}\right) \nonumber\\
&  -\mathit{E}_{1}\left(  {\frac{c}{b+1}}\right)  c+\exp\left(  -\frac{c}%
{b+1}\right)  b, \label{I1}%
\end{align}
where $\mathit{E}_{1}\left(  x\right)  $ is the exponential integral and where
we have used the definition $\left(  \ref{bc}\right)  $. Since $b\gg1$, we can
gain more information by expanding $\left(  \ref{I1}\right)  $ in terms of
$b$. We find%
\begin{equation}
b-c-e{^{-c}}+1+\mathit{E}_{1}\left(  c\right)  c+\gamma c{+c\ln}\left(
c/b\right)  +\mathcal{O}\left(  \frac{1}{b}\right)  . \label{I1a}%
\end{equation}
As regards the second integral of $P_{\mathrm{V}}$ of the r.h.s. of
Eq.$\left(  \ref{Pv1}\right)  $, we find%
\begin{align}
&  E_{\mathrm{Pl}}\int_{b}^{+\infty}\exp\left(  -c\left(  \alpha x^{2}-\beta
x\right)  \right)  dx\nonumber\\
&  =-E_{\mathrm{Pl}}{\frac{\sqrt{\pi}e{^{c\beta^{2}/4\alpha\,}}\left(
\operatorname{erf}{\left(  \sqrt{c}\left(  \alpha b-\beta/2\right)  \right)
}-1\right)  }{2\sqrt{c\alpha}}.} \label{I2}%
\end{align}
Combining $\left(  \ref{I1}\right)  $ with $\left(  \ref{I2}\right)  $ and
taking the leading order in $b$, we get%
\begin{equation}
P_{\mathrm{V}}=\frac{\exp\left(  cb(\beta-\alpha b)\right)  }{2c\alpha b^{2}%
}~.
\end{equation}

%%%%%%%%%%%%%%%%%%%%%%%%%%%%%%%%%%%%%%%%%%%%%%%%


\begin{thebibliography}{99}                                                                                               %
%\cite{Starobinsky:1980te}


\bibitem {Starobinsky:1980te}A.~A.~Starobinsky,
%``A New Type of Isotropic Cosmological Models Without Singularity,''
Phys.\ Lett.\ B \textbf{91} (1980) 99;
%%CITATION = PHLTA,B91,99;%%
%\cite{Guth:1980zm}
%\bibitem{Guth:1980zm}
A.~H.~Guth,
%``The Inflationary Universe: A Possible Solution to the Horizon and Flatness Problems,''
Phys.\ Rev.\ D \textbf{23} (1981) 347;
%%CITATION = PHRVA,D23,347;%%
%\cite{Linde:1981mu}
%\bibitem{Linde:1981mu}
A.~D.~Linde,
%``A New Inflationary Universe Scenario: A Possible Solution of the Horizon, Flatness, Homogeneity, Isotropy and Primordial Monopole Problems,''
Phys.\ Lett.\ B \textbf{108} (1982) 389.
%%CITATION = PHLTA,B108,389;%%


\bibitem {onset}E.\ Calzetta and M.\ Sakellariadou, Phys.\ Rev. \textbf{D 45}
(1992) 2802; E.\ Calzetta and M.\ Sakellariadou, Phys.\ Rev. \textbf{D 47}
(1993) 3184; C.\ Germani, W.\ Nelson and M.\ Sakellariadou, Phys.\ Rev.
\textbf{D 76} (2007) 043529 [arXiv:gr-qc/0701172].

%\cite{Gottlober:1990um}


\bibitem {Gottlober:1990um}S.~Gottlober, V.~Muller and A.~A.~Starobinsky,
%``Analysis of inflation driven by a scalar field and a curvature squared term,''
Phys.\ Rev.\ D \textbf{43} (1991) 2510.
%%CITATION = PHRVA,D43,2510;%%


%\cite{Bezrukov:2007ep}


\bibitem {Bezrukov:2007ep}F.~L.~Bezrukov and M.~Shaposhnikov,
%``The Standard Model Higgs boson as the inflaton,''
Phys.\ Lett.\ B \textbf{659} (2008) 703 [arXiv:0710.3755 [hep-th]];
%%CITATION = ARXIV:0710.3755;%%
%\cite{Barbon:2009ya}
%\bibitem{Barbon:2009ya}
J.~L.~F.~Barbon and J.~R.~Espinosa,
%``On the Naturalness of Higgs Inflation,''
Phys.\ Rev.\ D \textbf{79} (2009) 081302 [arXiv:0903.0355 [hep-ph]];
%%CITATION = ARXIV:0903.0355;%%
%\cite{Buck:2010sv}
%\bibitem{Buck:2010sv}
M.~Buck, M.~Fairbairn and M.~Sakellariadou,
%``Inflation in models with Conformally Coupled Scalar fields: An application to the Noncommutative Spectral Action,''
Phys.\ Rev.\ D \textbf{82} (2010) 043509 [arXiv:1005.1188 [hep-th]];
%%CITATION = ARXIV:1005.1188;%%
%\cite{Atkins:2010yg}
%\bibitem{Atkins:2010yg}
M.~Atkins and X.~Calmet,
%``Remarks on Higgs Inflation,''
Phys.\ Lett.\ B \textbf{697} (2011) 37 [arXiv:1011.4179 [hep-ph]];
%%CITATION = ARXIV:1011.4179;%%
%\cite{Cacciapaglia:2013tga}
%\bibitem{Cacciapaglia:2013tga}
G.~Cacciapaglia and M.~Sakellariadou,
%``Is F-term hybrid inflation natural within minimal supersymmetric SO(10)?,''
arXiv:1306.3242 [hep-ph].

\bibitem {MagSmo}J. Magueijo and L. Smolin, Class.\ Quant.\ Grav.
\textbf{21}(2004) 1725 [arXiv:gr-qc/0305055].

\bibitem {DSR}J.\ Kowalski-Glikman, \textsl{Phys. Lett. A} \textbf{286} (2001)
391 [arXiv:hep-th/0102098]; N.R. Bruno, G. Amelino-Camelia, J.
Kowalski-Glikman, Phys.\ Lett. \textbf{522} (2001) 133; G.~ Amelino-Camelia,
Int.\ J.\ Mod.\ Phys. \textbf{D 11} (2002) 35 [arXiv:gr-qc/0012051];
G.\ Amelino-Camelia, Phys.\ Lett. \textbf{B 510} (2001) 255
[arXiv:hep-th/0012238]; J.\ Magueijo and L.\ Smolin,
Phys.\ Rev.\ Lett.\ \textbf{88} (2002) 190403.

\bibitem {Vilenkin}A.\ Vilenkin, Phys.\ Rev. \textbf{D 37} (1988) 888.

\bibitem {Wiltshire}D.\ L.\ Wiltshire, \textsl{An introduction to quantum
cosmology} [arXiv:gr-qc/0101003]; D.\ L.\ Wiltshire, Gen.\ Rel.\ Grav.
\textbf{32} (2000) 515 [arXiv: gr-qc/9905090]; N.\ Kontoleon and
D.\ L.\ Wiltshire, Phys.\ Rev. \textbf{D 59} (1999) 063513 [arXiv:gr-qc/9807075].

\bibitem {HH}J.\ B.\ Hartle and S.\ W.\ Hawking, Phys.\ Rev. \textbf{D 28}
(1983) 2960; S.\ W.\ Hawking, Nucl.\ Phys. \textbf{B 239} (1984) 257.

\bibitem {LGGG}G.\ W.\ Gibbons and L.\ P.\ Grishchuk, Nucl.\ Phys. \textbf{B
313} (1989) 736.

\bibitem {WPKA}S.\ Watson, M.\ J.\ Perry, G.\ L.\ Kane and F.\ C.\ Adams, JCAP
0711 017 (2007) [arXiv: hep-th/0610054].

\bibitem {AM}S.\ Alexander and J.\ Magueijo, \textsl{Noncommutative geometry
as a realization of varying speed of light cosmology} in Proceedings of the
XIIIrd Rencontres de Blois, pp281, 2004 [arXiv:hep-th/0104093].

\bibitem {GaMaLo}R.\ Garattini, Phys.\ Lett. \textbf{B} 685 (2010) 329
[arXiv:0902.3927 [gr-qc]]; R.\ Garattini and F.\ S.\ N.\ Lobo, Phys.\ Rev.
\textbf{D 85} (2012) 024043 [arXiv:1111.5729 [gr-qc]]; R.\ Garattini and
G.\ Mandanici, Phys.\ Rev. \textbf{D 83} (2011) 084021 [arXiv:1102.3803
[gr-qc]]; R.\ Garattini, JCAP 1306 (2013) 017 [arXiv:1210.7760 [gr-qc]].

\bibitem {GaMa}R.\ Garattini and G.\ Mandanici, Phys.\ Rev. \textbf{D 85}
(2012) 023507 [arXiv:1109.6563 [gr-qc]].

\bibitem {CM}D.\ H.\ Coule and J.\ Martin, Phys.\ Rev. \textbf{D 61} (2000)
063501 [arXiv:gr-qc/9905056].
\end{thebibliography}
\end{document}